\documentclass[runningheads]{llncs}
\usepackage{graphicx}
\usepackage{multirow}
\usepackage[table,xcdraw]{xcolor}
\usepackage{subfig}

\usepackage{tabu}  

\begin{document}
\title{CobotTouch: AR-based Interface with Fingertip-worn Tactile Display for Immersive Operation/Control of Collaborative Robots } 
\titlerunning{CobotTouch}
%
\author{Oleg Sautenkov\inst{1} \and 
Miguel Altamirano Cabrera\inst{1} \and 
Viktor Rakhmatulin\inst{1} \and 
Dzmitry Tsetserukou\inst{1}
}
\authorrunning{O. Sautenkov et al.}
%
\institute{Center for Digital Engineering, Skolkovo Institute of Science and Technology (Skoltech), 121205 Bolshoy Boulevard 30, bld. 1, Moscow, Russia.
\email{\{oleg.sautenkov, miguel.altamirano, viktor.rakhmatulin, d.tsetserukou\}@skoltech.ru}}
\maketitle              
\begin{abstract}
Complex robotic tasks require human collaboration to benefit from their high dexterity. Frequent human-robot interaction is mentally demanding and time-consuming. Intuitive and easy-to-use robot control interfaces reduce the negative influence on workers, especially inexperienced users. In this paper, we present CobotTouch, a novel intuitive robot control interface with fingertip haptic feedback. The proposed interface consists of a projected Graphical User Interface on the robotic arm to control the position of the robot end-effector based on gesture recognition, and a wearable haptic interface to deliver tactile feedback on the user's fingertips. We evaluated the user's perception of the designed tactile patterns presented by the haptic interface and the intuitiveness of the proposed system for robot control in a use case. The results revealed a high average recognition rate of 75.25\% for the tactile patterns. An average NASA Task Load Index (TLX) indicated small mental and temporal demands proving a high level of the intuitiveness of CobotTouch for interaction with collaborative robots.%

\end{abstract}
\section{Introduction}

Industrial collaborative robots are becoming more important in the modern manufacturing industry. Collaborative robots perform routine and repeatable tasks while workers are focused on dexterous operations. Robot operations are complex and require expert knowledge in the robotics domain. Moreover, changes in the production in Small and Medium Enterprises (SMEs) are frequent and lead to an increase in the time that workers interact with robotic systems \cite{SME_robotics}. Intuitive and efficient control interfaces significantly decrease the number of errors produced by operators \cite{7819154}. 

Virtual Reality (VR) technologies are widely used in entertainment, education, and other business applications. 
Augmented Reality (AR) and VR provide alternative robot control interfaces to traditional ones, such as teaching pendants and joysticks. The technologies are aimed to support efficient and low cognitive demanding human-robot interaction \cite{KagermannWahlsterHelbig2013en}. For example, C. dos Santos et. al. \cite{VREducation} designed a 3D environment to teach children to program robots using a head-mounted display (HMD). VR technologies can be used for sharing tasks between workers and an industrial robot naturally and intuitively, as presented by Beibei et. al.  \cite{VRTaskSharing}. However, VR devices have numerous limitations, such as restricted field-of-view and motion sickness \cite{VRSickness}.

\begin{figure}[t!]
  \centering
  \includegraphics[width=0.45\linewidth]{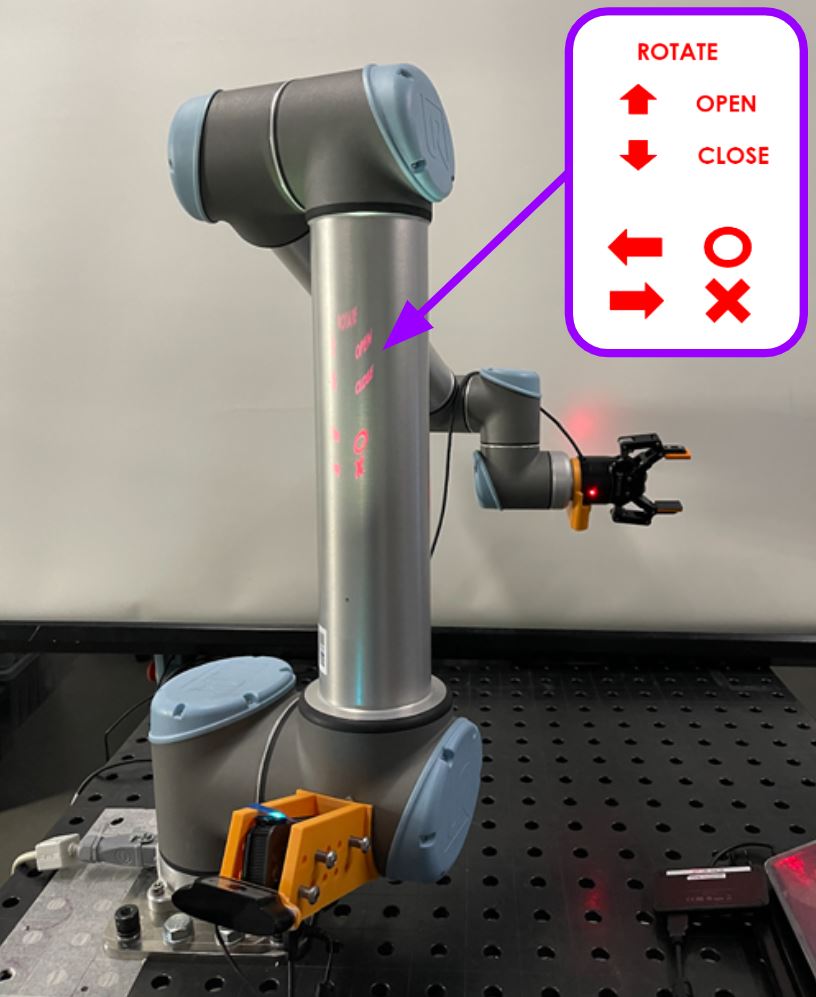}
  \caption{Projected Graphical User Interface on the robotic arm with DNN-based hand gesture recognition to control the position of a collaborative robot UR10 by CobotTouch.}
  \vspace{-0.7cm}
  \label{fig:SystemUR3}
\end{figure}

AR enriches the surrounding environment by virtual layers, providing the user with the required information and enabling interacting interfaces. AR has found a variety of applications in Human-Robot Interaction (HRI). A complete setup for safe and efficient HRI was presented by Papanastasiou et al. \cite{Papanastasiou2019}. Smart glasses provide the AR interface to visualize safety zones. However, the projection mapping approach overcomes limitations in a restricted field of view and motion sickness. Projected GUIs allow interaction with the virtual control panel at almost any type of surface as shown in \cite{10.1145/1667146.1667160}. Hartmann et al. \cite{10.1145/3379337.3415849} presented a combined system with a head-mounted projector and a Hololens AR headset that allow multi-user collaboration and shared use of the projected interface.

VR and AR-based technologies rely mostly on visual modality, while in cases with limited or absent visual feedback, tactile sense could complement HRI and make it more intuitive. Wearable haptic interfaces deliver the sense of grasping for solid and deformable bodies as presented in \cite{Haptic_GrabityWeight} and in \cite{hapticForSpatialTasks}, slippage of manipulated virtual objects \cite{FingertipTanForce}, and other interactive environment parameters that are hardly identifiable by visual channel. The tactile stimuli are often applied on the palm or the fingertips because of the high skin sensitivity in these parts of the body, and because usually are used to interact with the surroundings of the environment. As shown by Gabardi et al. \cite{hapticsFingertipShapes}, a finger-worn haptic device could successfully render the texture, curvature, edges, and orientation of an explored virtual surface.  

We present CobotTouch, a novel AR robot control interface that consists of a Camera-Projector Module (CPM), hand gesture recognition enabled by Deep Neural Network (DNN), and a tactile haptic interface for the fingertips. CobotTouch inherits the benefits of projected GUIs and tactile feedback, while there are no drawbacks typically caused by VR helmets. The CPM is mounted statically on the robot's body and projects a Graphical User Interface (GUI) to control the robot directly on the robotic link surface, as shown in Fig. \ref{fig:SystemUR3}. Gesture recognition captures the position of the hand while interacting with the GUI and allows dexterous manipulation of the robot's end-effector. Two-finger haptic interfaces guide the user during the control of the end-effector, rendering the orientation of the robotic gripper on the fingertips.

The remainder of the paper is organized as follows. In Section II, a detailed overview of the system is presented. In Section III a tactile pattern perception experiment was conducted to evaluate the users' responses to the haptic interface. Section IV describes a robot control use case where users evaluated the convenience and intuitiveness of the developed system. In the last section, the obtained results are discussed and further works are proposed.

\section{System Overview}

CobotTouch is a novel HRI system that implements AR projector-based spatial displays, which generates an interactive projected GUI on the links of a collaborative robot UR10. The system tracks the position of the user's hand using DNN-based gesture recognition. It allows users to interact with the projected GUI and control the robot's position intuitively. The projector-camera module is mounted statically on the shoulder of the robot. Two wearable finger haptic interfaces, introduced in \cite{linkRing}, guide the user during the control of the end-effector, rendering the orientation of the robotic gripper on the fingertips.

The hardware components of the CobotTouch system are a pico pocket projector Optoma PK301, a Logitech HD Webcam C930e, a 6 DoF collaborative robot from Universal Robots UR10 \cite{UR10}, a two-finger gripper from Robotiq 2f-85 \cite{Robotiq}, two haptic interfaces, and a laptop.
 
\begin{figure}[h!]
  \centering
  \includegraphics[width=0.6\linewidth]{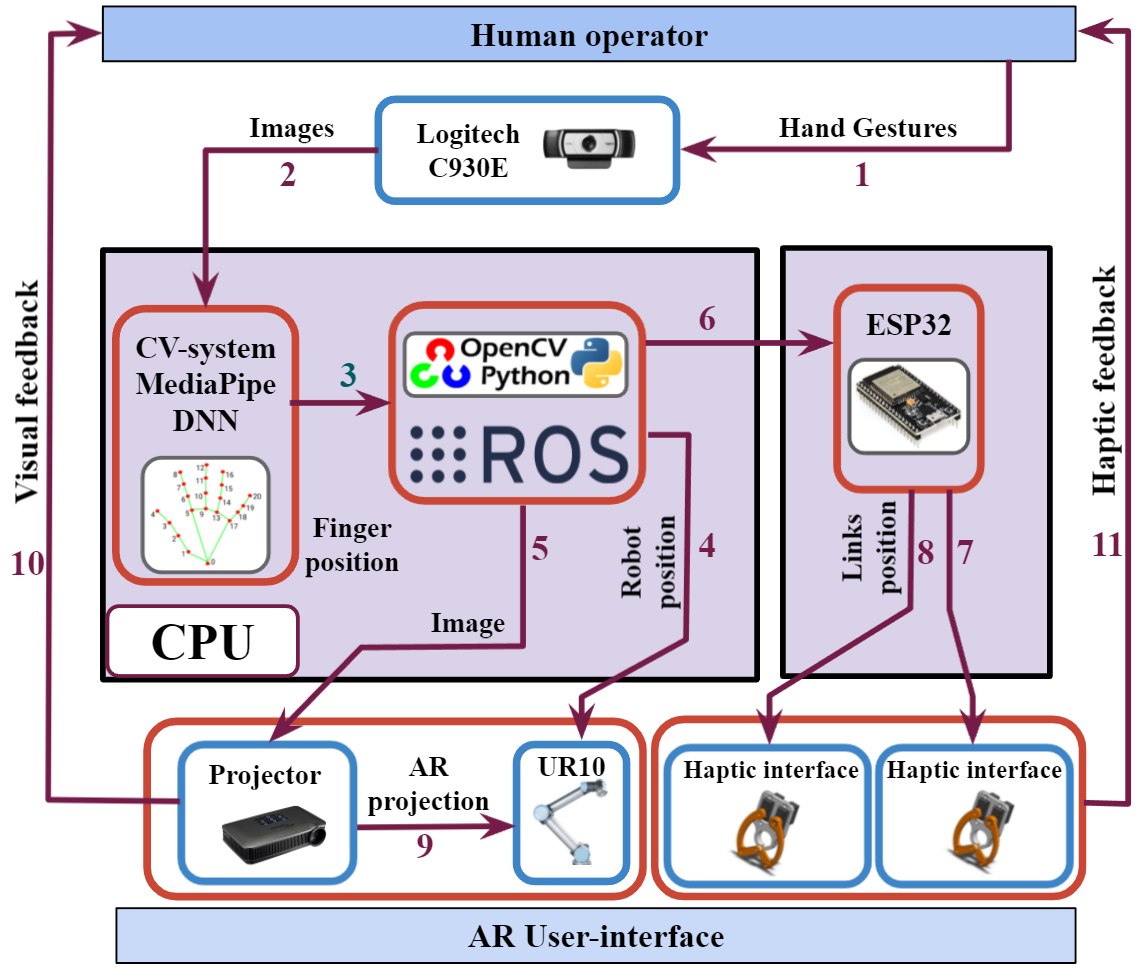}
  \caption{System architecture.}
  \vspace{-0.3cm}
  \label{fig:System architecture}
\end{figure}

\begin{figure}[t!]
  \centering
  \includegraphics[width=0.6\linewidth]{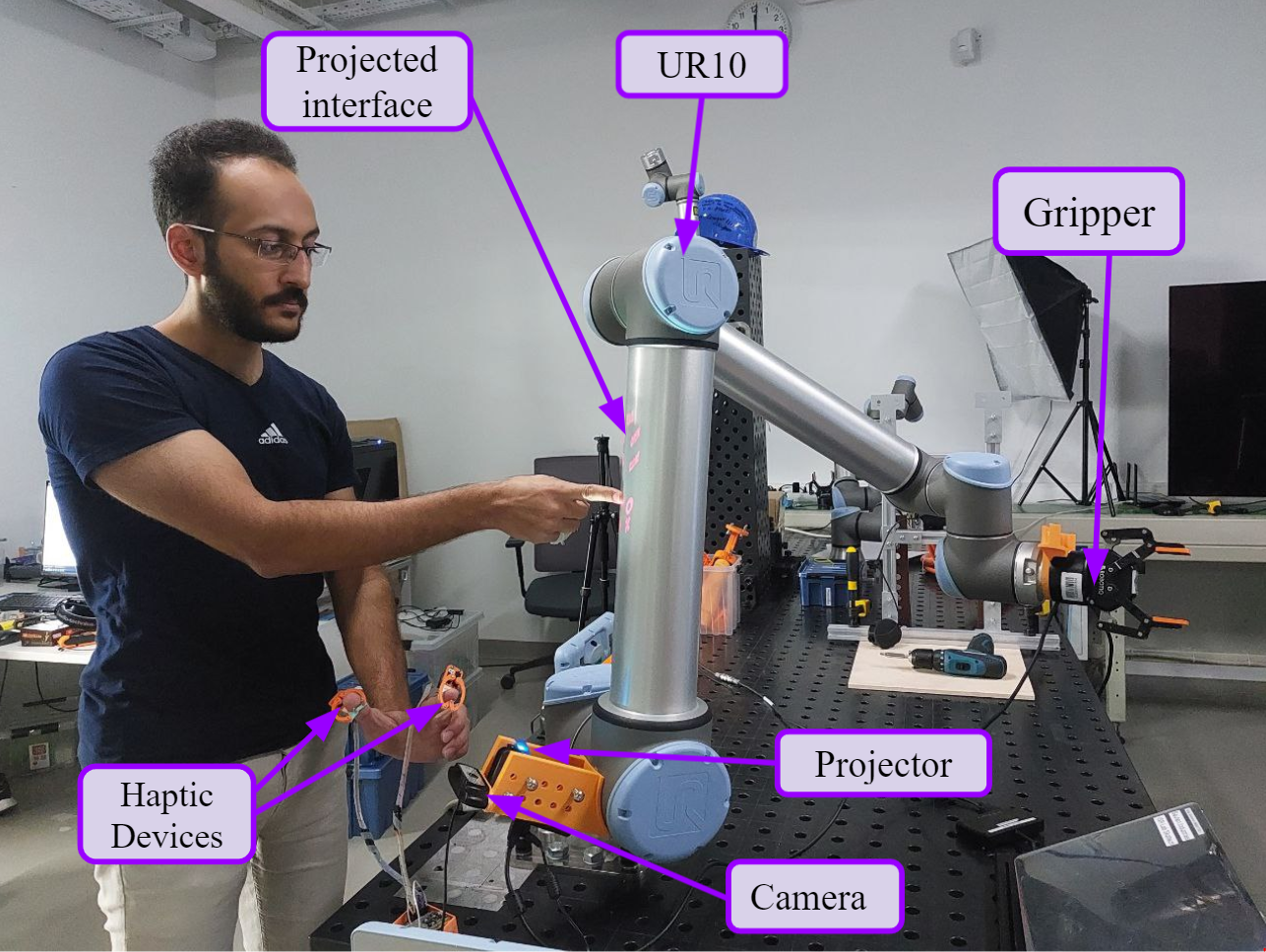}
  \caption{User controls the UR10 collaborative robot through the projected interface from ultra-compact mobile projector and  HD Webcam C930e mounted on the joint of the UR10 robot.}
  \vspace{-0.5cm}
  \label{fig:System overview}
\end{figure}

Three computational modules are responsible for data processing and robot control: a) gesture recognition based on DNN, b) image processing through the OpenCV library for the projection, and c) haptic interface control. The system architecture is shown in Fig. \ref{fig:System architecture}, and the system overview is shown in Fig. \ref{fig:System overview}. The software architecture is based on the ROS Melodic framework.

The DNN-based algorithm detects the user's hand gestures. It allows users to interact with any type of GUI projection and control the robot's position more intuitively. The CPM is mounted on the UR10 robot to provide a more extensive projection area and to avoid shadows in the projection. Simultaneously, the Computer Vision (CV) algorithm processes the Webcam image using the hand tracking module. It estimates the position of the fingers in a specific area and defines it as a ``press button" command. As a result of the gesture recognition process, the central program defines the required robotic action. CobotTouch can present different GUIs with buttons and interact with the robotic parts, e.g. users can visualize the inner structure of the robot by projecting on the robot without disassembling it.

\subsection{Haptic Interface}

\begin{figure}[!thb]
  \centering%
  \vspace{-0.4cm}
  \subfloat[3D CAD model of wearable tactile display LinkTouch.]{%
         \label{fig:device}%
         \includegraphics[width=0.46\linewidth]{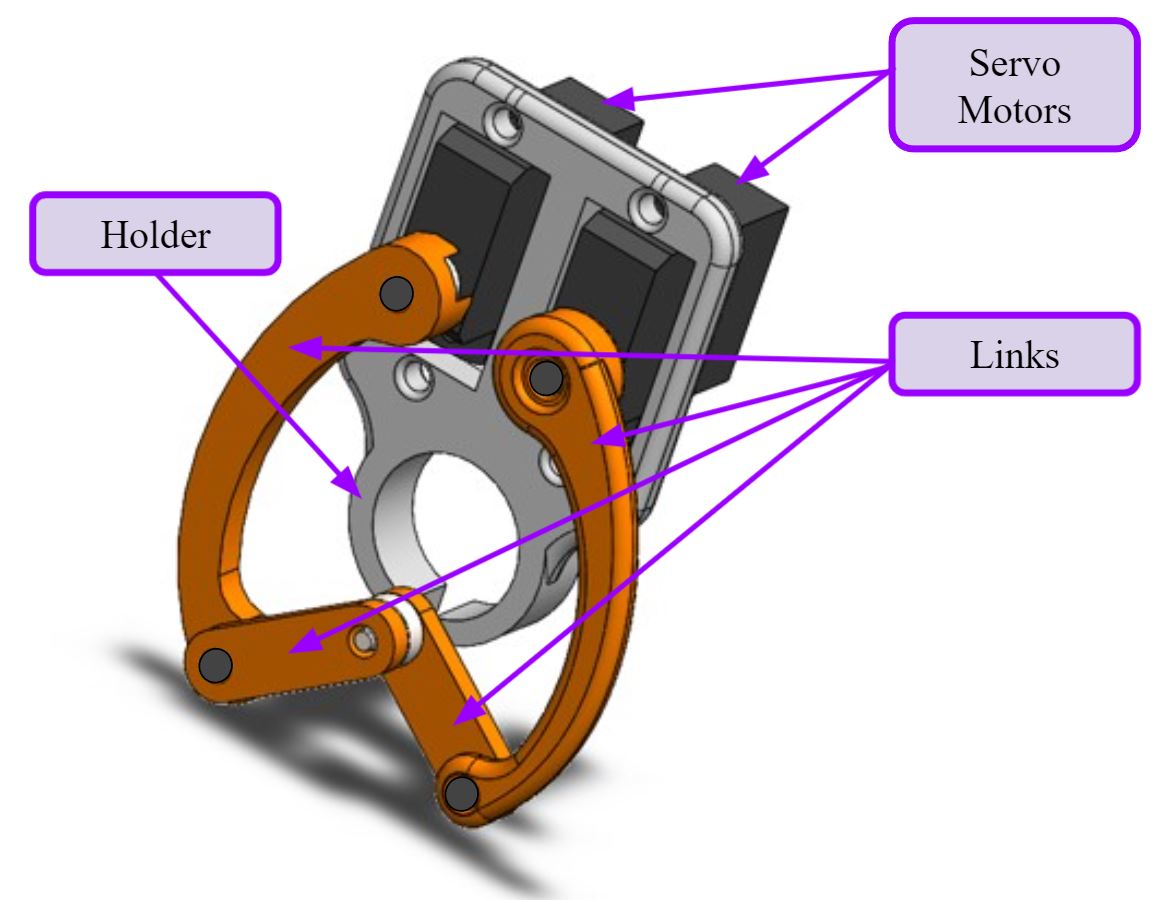}%
} \qquad
\subfloat[Example of rotational patterns. The purple arrows correspond to the counterclockwise pattern, and the yellow arrows correspond to the clockwise pattern.]{%
      \label{fig:Rotational patterns}%
      \includegraphics[width=0.46\linewidth]{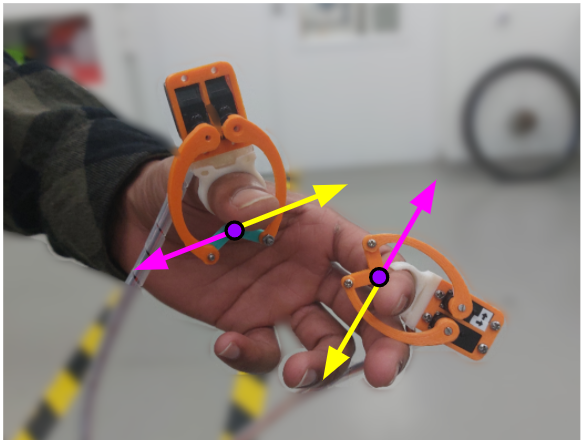}%
    }

  \caption{Wereable tactile display LinkTouch, used to provide tactile feedback on the thumb and index fingertips.}
  \vspace{-0.6cm}
   \label{fig:typical_patterns}
\end{figure}

We used two devices with a five-bar linkage mechanism as a tactile feedback module on the thumb and index fingertips. The devices are based on LinkTouch \cite{6775473} and LinkRing \cite{linkRing} technology. The device consists of a 3D printed PLA body, a 3D printed flexible material finger cap holder, links, and two DSM44 servo motors, as shown in Fig. \ref{fig:device}. 
Two ESP32 microcontrollers are used to control the interfaces. The program sends a signal to the microcontrollers of the devices when a trigger occurs, and the haptic devices become active, generating normal force on the desired position of the fingertips.

The haptic interface allows the user to interact more efficiently with the robot by getting haptic feedback support at his fingertips. The configuration of the five bar-linkage mechanisms allows one to perceive the position changing of the joint on the fingertip. By combining the two LinkTouch devices, we generated rotational patterns, as shown in Fig. \ref{fig:Rotational patterns}. Wearing two Linktouch haptic interfaces, the operator can understand the direction of the robot's rotation while grasping objects, get hints for the next movement, or feel the gripper's direction.

\subsection{DNN-based Gesture Recognition}

The DNN-based gesture recognition module is implemented based on the Mediapipe framework \cite{zhang2020mediapipe}. It provides high-fidelity hand tracking by employing Machine Learning (ML) to infer  21 3D landmarks of a human hand per frame.

The DNN algorithm was trained to recognize eight gestures. Two gestures were chosen to perform the pressing buttons task, one with the open fingers (``Palm") and the second with only the index finger pointing (``One").  

If the index finger coordinates are located on the button's area, and the gesture has been changed from ``Palm" to ``One," the corresponding button will activate. The algorithm sends the number of the active buttons by ROS framework to the robot control system. The control system modifies the position of the robot end-effector during the time that the button has been pressed, and the projected GUI changes the color of the pressed button to inform the user that the action is in progress.

\section{Experiment on Tactile Perception}

\begin{figure}[!thb]
  \centering%
  \vspace{-0.3cm}
  \subfloat[Set of tactile patterns represented on the thumb and index fingertips, the arrows represent the sliding direction by the haptic interface contact point.]{%
         \label{fig:Patterns}%
         \includegraphics[width=0.46\linewidth]{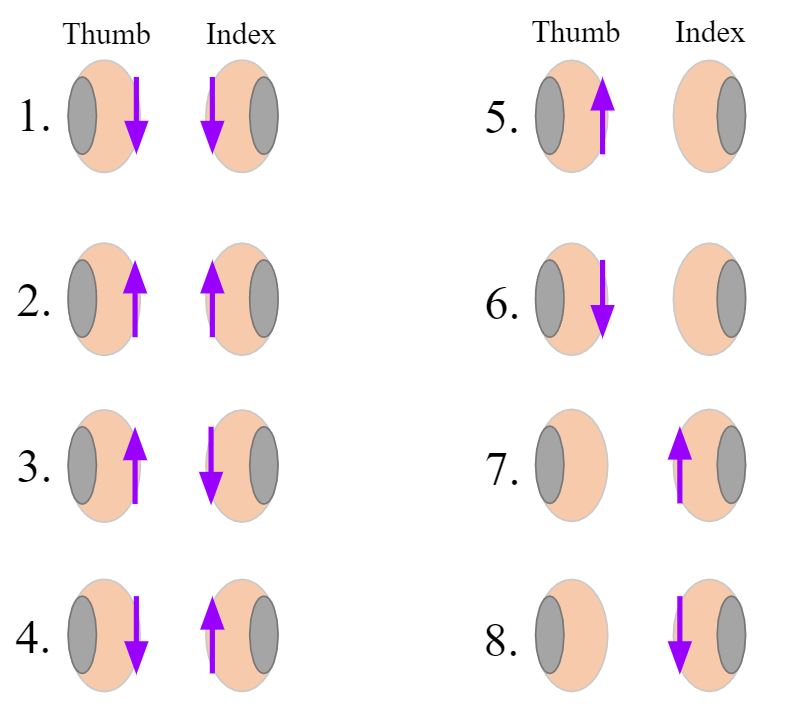}%
} \qquad
\subfloat[Patterns perception experimental setup.]{%
      \label{fig:PatternsSetup}%
      \includegraphics[width=0.46\linewidth]{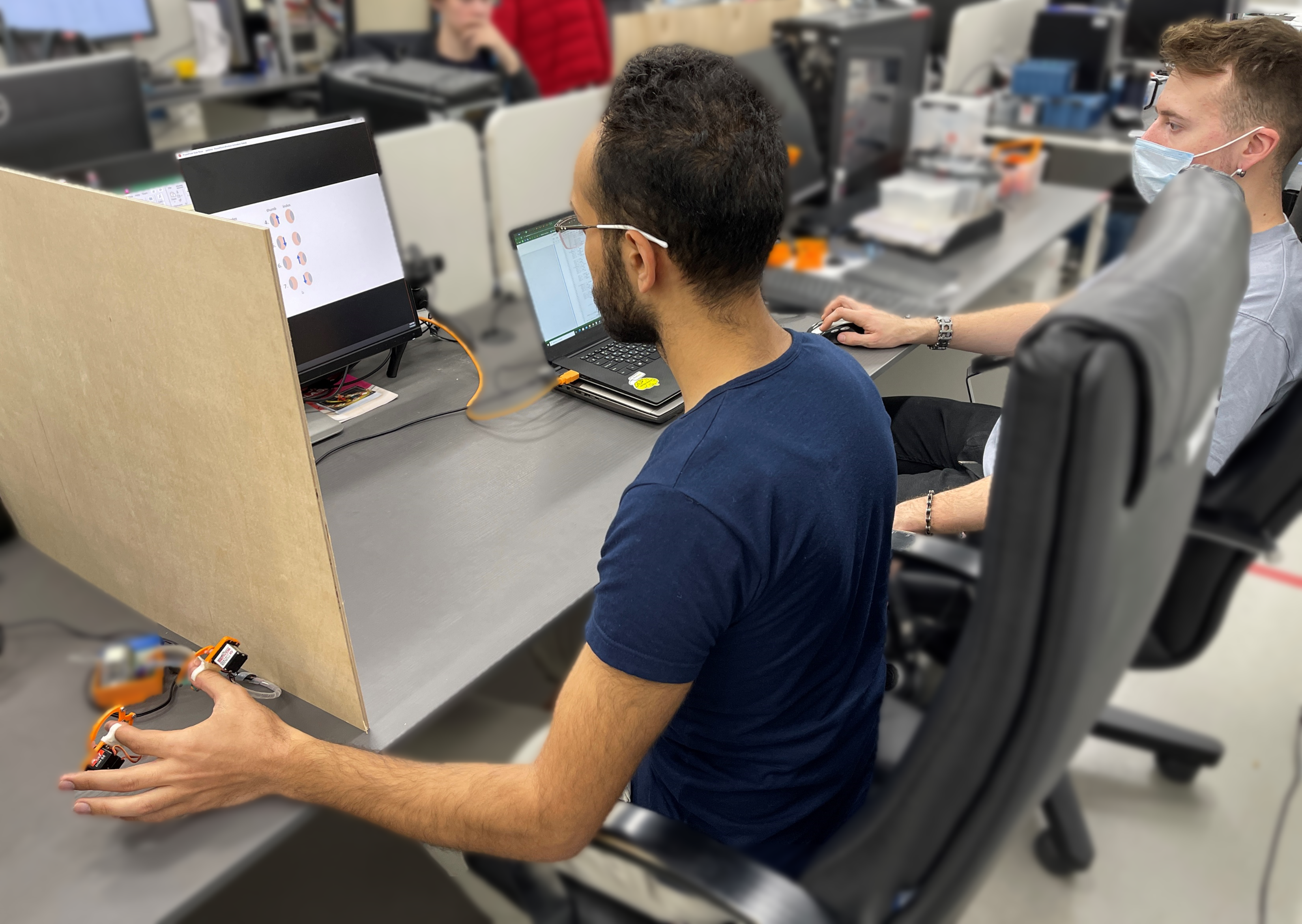}%
    }
    
 \caption{Experiment on tactile perception.}
  \vspace{-0.5cm}
   \label{fig:typical_patterns2}
\end{figure}

This evaluation is centered on the analysis of the human perception of the tactile rendering on the fingertips. Eight patterns were designed to evaluate the human perception when the thumb and index fingertips were stimulated simultaneously or once per time. The contact points of the haptic interface slide on the fingertips in different directions according to the patterns as shown in Fig. \ref{fig:Patterns}.

Seven right-handed participants (2 females) aged 22 to 32 years volunteering completed the evaluation. None of them reported any deficiencies in sensorimotor function.   

Before the experiment, the device was calibrated, and a training session was performed. During the training session,  the experimenter explained the purpose of the haptic device to each participant and demonstrated the tactile patterns at least three times.
During the experiment, the user was asked to sit in front of a desk and to wear the haptic display on the left thumb and index fingers as shown in Fig. \ref{fig:PatternsSetup}. A visual barrier was located between the left hand and the user. On the screen, all the possible available patterns were displayed on the screen during the experiment. The users were asked to tell the perceived pattern to the experimenter, who recorded the pattern and time. Each pattern was presented five times blindly in random order, thus, 40 patterns were provided to each participant in each evaluation.

\subsection{Tactile Perception Results}

The results of the human pattern perception are summarized in the confusion matrix shown in Table I.

\begin{table}
\vspace{-0.5cm}
\caption{Confusion Matrix for Actual and Perceived
Pattern Recognition Across All Subjects.}
\label{CM3}
\setlength{\tabcolsep}{2pt} 
\renewcommand{\arraystretch}{1} 
\centering{

\begin{tabular}{|cc|cccccccc|}
\hline

\multicolumn{2}{|c|}{} &
  \multicolumn{8}{c|}{\textit{Answers   (Predicted Class)}} \\ \cline{3-10} 
\multicolumn{2}{|c|}{\multirow{-2}{*}{\%}} &
  \multicolumn{1}{c|}{1} &
  \multicolumn{1}{c|}{2} &
  \multicolumn{1}{c|}{3} &
  \multicolumn{1}{c|}{4} &
  \multicolumn{1}{c|}{5} &
  \multicolumn{1}{c|}{6} &
  \multicolumn{1}{c|}{7} &
  8 \\ \hline
\multicolumn{1}{|c|}{} &
  1 &
  \multicolumn{1}{c|}{\cellcolor[HTML]{5976AB}{\color[HTML]{F9FAFB} 0.70}} &
  \multicolumn{1}{c|}{\cellcolor[HTML]{E5E9F3}0.10} &
  \multicolumn{1}{c|}{\cellcolor[HTML]{DAE0ED}0.15} &
  \multicolumn{1}{c|}{\cellcolor[HTML]{F1F3F9}0.05} &
  \multicolumn{1}{c|}{\cellcolor[HTML]{FCFCFF}0.00} &
  \multicolumn{1}{c|}{\cellcolor[HTML]{FCFCFF}0.00} &
  \multicolumn{1}{c|}{\cellcolor[HTML]{FCFCFF}0.00} &
  \cellcolor[HTML]{FCFCFF}0.00 \\ \cline{2-10} 
\multicolumn{1}{|c|}{} &
  2 &
  \multicolumn{1}{c|}{\cellcolor[HTML]{F1F3F9}0.05} &
  \multicolumn{1}{c|}{\cellcolor[HTML]{3C5E9C}{\color[HTML]{F9FAFB} 0.83}} &
  \multicolumn{1}{c|}{\cellcolor[HTML]{F1F3F9}0.05} &
  \multicolumn{1}{c|}{\cellcolor[HTML]{FCFCFF}0.00} &
  \multicolumn{1}{c|}{\cellcolor[HTML]{FCFCFF}0.00} &
  \multicolumn{1}{c|}{\cellcolor[HTML]{F7F8FC}0.03} &
  \multicolumn{1}{c|}{\cellcolor[HTML]{F7F8FC}0.03} &
  \cellcolor[HTML]{F7F8FC}0.03 \\ \cline{2-10} 
\multicolumn{1}{|c|}{} &
  3 &
  \multicolumn{1}{c|}{\cellcolor[HTML]{EBEEF6}0.08} &
  \multicolumn{1}{c|}{\cellcolor[HTML]{EBEEF6}0.08} &
  \multicolumn{1}{c|}{\cellcolor[HTML]{7C93BD}{\color[HTML]{F9FAFB} 0.55}} &
  \multicolumn{1}{c|}{\cellcolor[HTML]{BCC8DE}0.28} &
  \multicolumn{1}{c|}{\cellcolor[HTML]{F7F8FC}0.03} &
  \multicolumn{1}{c|}{\cellcolor[HTML]{FCFCFF}0.00} &
  \multicolumn{1}{c|}{\cellcolor[HTML]{FCFCFF}0.00} &
  \cellcolor[HTML]{FCFCFF}0.00 \\ \cline{2-10} 
\multicolumn{1}{|c|}{} &
  4 &
  \multicolumn{1}{c|}{\cellcolor[HTML]{F7F8FC}0.03} &
  \multicolumn{1}{c|}{\cellcolor[HTML]{F1F3F9}0.05} &
  \multicolumn{1}{c|}{\cellcolor[HTML]{E5E9F3}0.10} &
  \multicolumn{1}{c|}{\cellcolor[HTML]{4E6CA5}{\color[HTML]{F9FAFB} 0.75}} &
  \multicolumn{1}{c|}{\cellcolor[HTML]{EBEEF6}0.08} &
  \multicolumn{1}{c|}{\cellcolor[HTML]{FCFCFF}0.00} &
  \multicolumn{1}{c|}{\cellcolor[HTML]{FCFCFF}0.00} &
  \cellcolor[HTML]{FCFCFF}0.00 \\ \cline{2-10} 
\multicolumn{1}{|c|}{} &
  5 &
  \multicolumn{1}{c|}{\cellcolor[HTML]{F7F8FC}0.03} &
  \multicolumn{1}{c|}{\cellcolor[HTML]{FCFCFF}0.00} &
  \multicolumn{1}{c|}{\cellcolor[HTML]{FCFCFF}0.00} &
  \multicolumn{1}{c|}{\cellcolor[HTML]{F7F8FC}0.03} &
  \multicolumn{1}{c|}{\cellcolor[HTML]{42639F}{\color[HTML]{F9FAFB} 0.80}} &
  \multicolumn{1}{c|}{\cellcolor[HTML]{DFE4F0}0.13} &
  \multicolumn{1}{c|}{\cellcolor[HTML]{F7F8FC}0.03} &
  \cellcolor[HTML]{FCFCFF}0.00 \\ \cline{2-10} 
\multicolumn{1}{|c|}{} &
  6 &
  \multicolumn{1}{c|}{\cellcolor[HTML]{F7F8FC}0.03} &
  \multicolumn{1}{c|}{\cellcolor[HTML]{FCFCFF}0.00} &
  \multicolumn{1}{c|}{\cellcolor[HTML]{FCFCFF}0.00} &
  \multicolumn{1}{c|}{\cellcolor[HTML]{FCFCFF}0.00} &
  \multicolumn{1}{c|}{\cellcolor[HTML]{E5E9F3}0.10} &
  \multicolumn{1}{c|}{\cellcolor[HTML]{5371A8}{\color[HTML]{F9FAFB} 0.73}} &
  \multicolumn{1}{c|}{\cellcolor[HTML]{DFE4F0}0.13} &
  \cellcolor[HTML]{F7F8FC}0.03 \\ \cline{2-10} 
\multicolumn{1}{|c|}{} &
  7 &
  \multicolumn{1}{c|}{\cellcolor[HTML]{F1F3F9}0.05} &
  \multicolumn{1}{c|}{\cellcolor[HTML]{FCFCFF}0.00} &
  \multicolumn{1}{c|}{\cellcolor[HTML]{FCFCFF}0.00} &
  \multicolumn{1}{c|}{\cellcolor[HTML]{FCFCFF}0.00} &
  \multicolumn{1}{c|}{\cellcolor[HTML]{FCFCFF}0.00} &
  \multicolumn{1}{c|}{\cellcolor[HTML]{FCFCFF}0.00} &
  \multicolumn{1}{c|}{\cellcolor[HTML]{4868A2}{\color[HTML]{F9FAFB} 0.78}} &
  \cellcolor[HTML]{D4DBEA}0.18 \\ \cline{2-10} 
\multicolumn{1}{|c|}{\multirow{-8}{*}{\textit{\rotatebox{90}{Patterns}}}} &
  8 &
  \multicolumn{1}{c|}{\cellcolor[HTML]{F7F8FC}0.03} &
  \multicolumn{1}{c|}{\cellcolor[HTML]{FCFCFF}0.00} &
  \multicolumn{1}{c|}{\cellcolor[HTML]{FCFCFF}0.00} &
  \multicolumn{1}{c|}{\cellcolor[HTML]{FCFCFF}0.00} &
  \multicolumn{1}{c|}{\cellcolor[HTML]{FCFCFF}0.00} &
  \multicolumn{1}{c|}{\cellcolor[HTML]{FCFCFF}0.00} &
  \multicolumn{1}{c|}{\cellcolor[HTML]{E5E9F3}0.10} &
  \cellcolor[HTML]{305496}{\color[HTML]{F9FAFB} 0.88} \\ \hline
\end{tabular} }
\end{table}

In order to evaluate the statistically significant differences between the pattern perception, we analyzed the results using single factor repeated-measures ANOVA, with a chosen significance level of $\alpha<0.05$. The sphericity and normality assumptions were examined and no violations were detected. According to the ANOVA results, there is a statistically significant difference in pattern perception, which corresponds to the patterns in Fig. \ref{fig:Patterns}, $F(7,48) = 2.077, p = 0.064$. The ANOVA results show that the pattern influences the percentage of correct responses. The average recognition rate of the patterns is 75\%. 

The paired t-tests showed statistically significant differences between the pattern 1 and 3 ($p=0.037 < 0.05$), 2 and 3 ($p=0.001 < 0.05$), 3 and 5 ($p=0.028 < 0.05$), 3 and 7 ($p=0.015 < 0.05$), and 3 and 8 ($p=0.026 < 0.05$). The open-source statistical package Pingouin \cite{Vallat2018PingouinSI} was used for the statistical analysis.


\section{User Study Experiment}
The principal approach of this paper is to design a new AR control interface with fingertip haptic feedback. We conducted a user study to investigate the advantages and disadvantages of the proposed system according to NASA Task Load Index (TLX) Rating. 

\subsection{Experimental Design}

\begin{figure}[ht!]
\centering
  \vspace{-0.3cm}
  \includegraphics[width=0.55\linewidth]{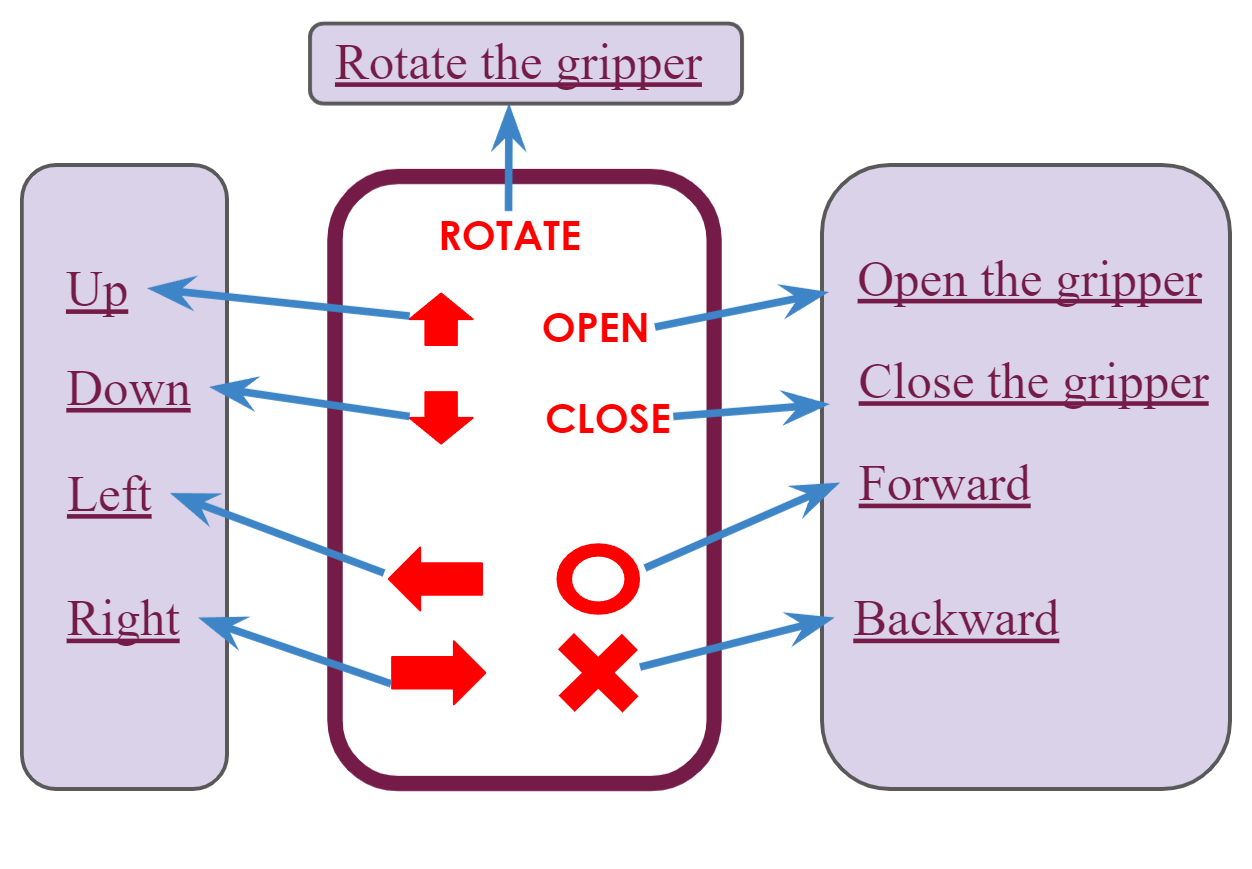}
  \caption{CobotTouch control interface. The interface contains nine buttons. The gesture recognition system detects the pressed button and moves the robot in the desired direction or perform pre-defined action.}
  \vspace{-0.5cm}
  \label{fig:CobotAR system interface}
\end{figure}

\begin{figure}[h!]
  \centering
  \includegraphics[width=0.7\linewidth]{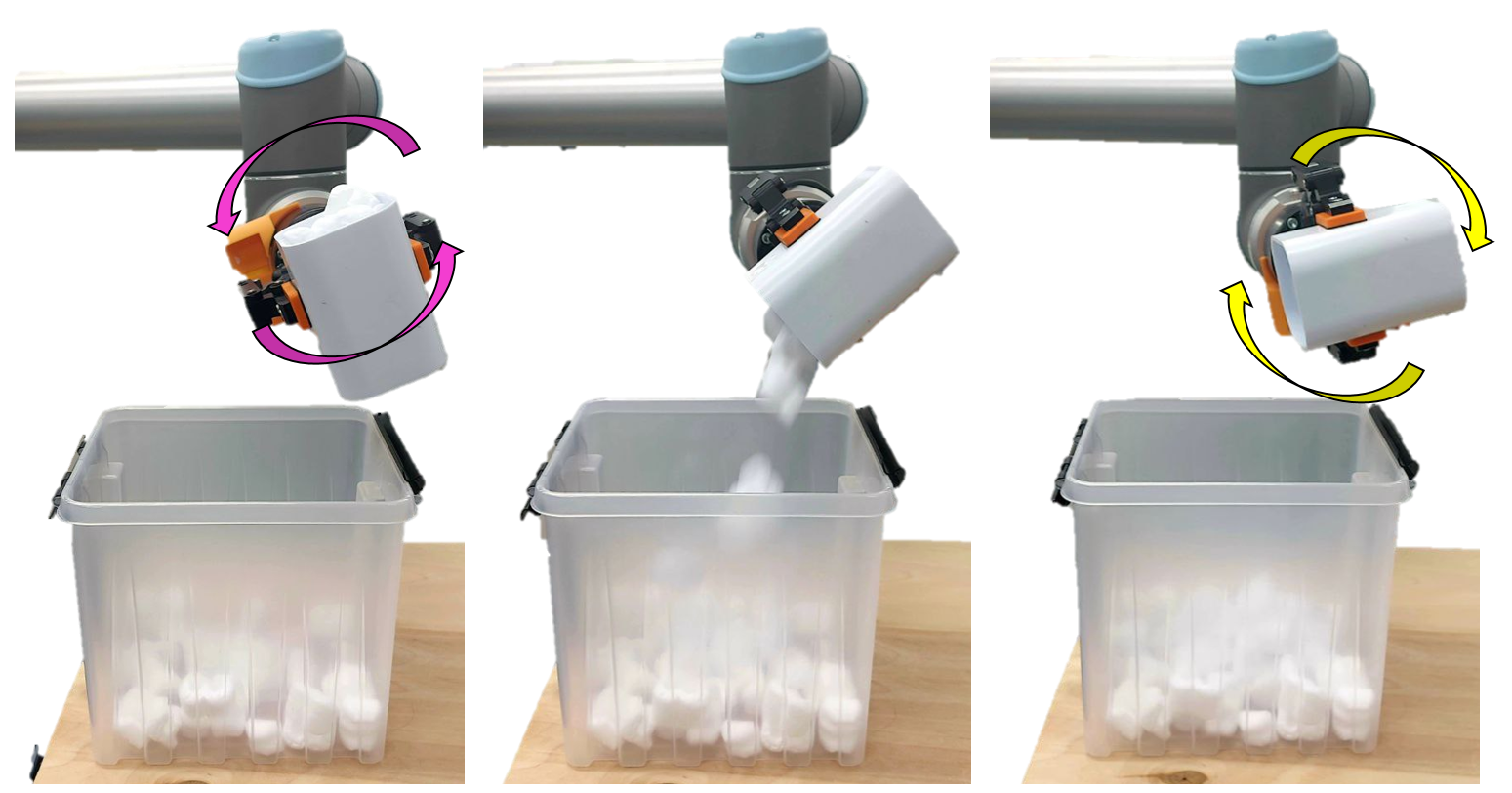}
  \caption{Gripper rotation during the user study experiment.}
  \vspace{-0.7cm}
  \label{fig:Gripper rotation}
\end{figure}

The GUI to control the UR10 robot through the CobotTouch system is represented in Fig. \ref{fig:CobotAR system interface}. By pressing on the projected buttons, the robot end-effector moves on the three axes. The user can use six different movement control buttons to move the robot. The button ``close" closes the gripper, the button ``open" opens it. To rotate the end-effector and to overturn the container, users should press the button ``rotate". When the gripper is rotating, the CPU constantly sends the rotational pattern to the haptic device as shown in Fig. \ref{fig:Rotational patterns} according to directions represented in Fig. \ref{fig:Gripper rotation}.

The user can use only one button simultaneously. During the experiment, the user can correct each time the robot trajectory. 

We located one empty plastic box and two containers filled with white styrofoam pieces on the experiment table as shown in Fig.\ref{fig:Rotational patterns}. Participants were asked to put the content of the two containers into the box by controlling the robot with the CobotTouch interface.

Before starting the experiment, the participants performed a training session, where each participant familiarized themselves with the interface and tested it. After the training session, the UR10 robot returned to the predetermined initial position, and the experiment started.

\begin{figure}[h!]
  \centering
  \includegraphics[width=0.6\linewidth]{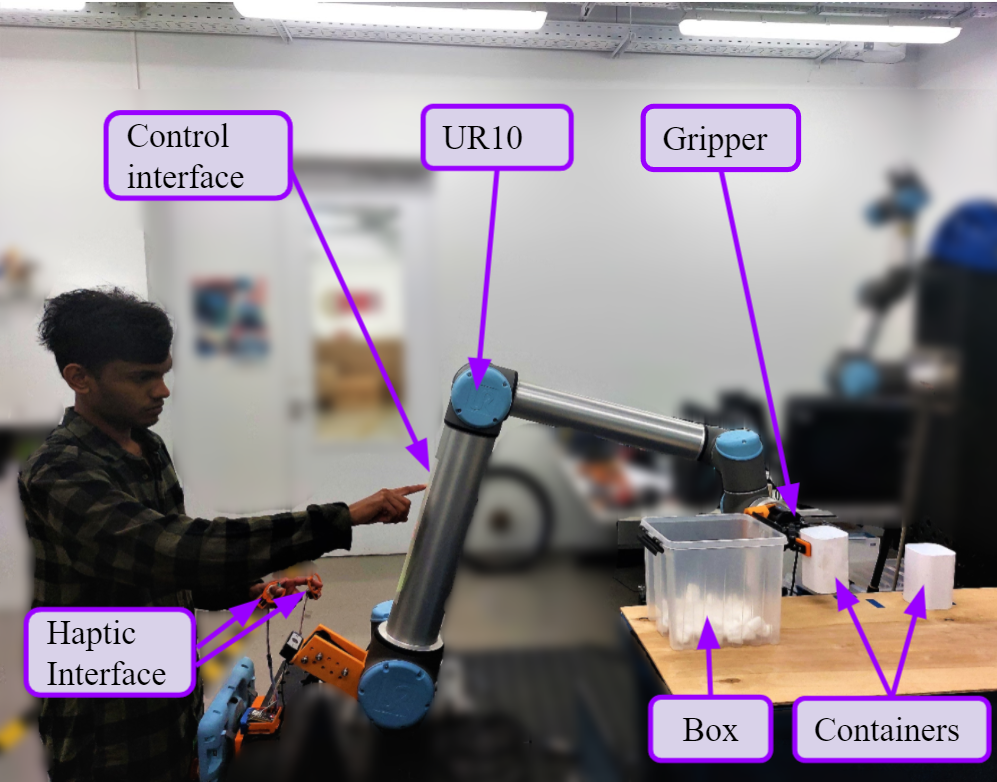}
  \caption{Setup for the user study experiment}
  \label{fig:CobotAR system controlling}
\end{figure}

After the trial, each participant completed a questionnaire based on The NASA TLX measuring the physical, mental, temporal, performance, effort, and frustration demands. The participants filled it in order to determine the advantages and disadvantages of using the CobotSystem system and gave some comments about their experience. Eight participants (2 females) volunteering conducted the test, aged from 22 to 32 years.

\section{Experimental Results}

\subsection{Average Score of NASA TLX Rating}

The average results of the NASA TLX ratings are shown in Table \ref{table_NASA}. It can be can observed that physical demand had the highest value of 2.08. One of the participants noticed that the performing of this task through the CobotTouch interface was going with a constantly raised hand, as shown in Fig. \ref{fig:CobotAR system controlling}, which could affect the evaluation of this parameter.

\begin{figure}[h!]
  \centering
  \includegraphics[width=0.6\linewidth]{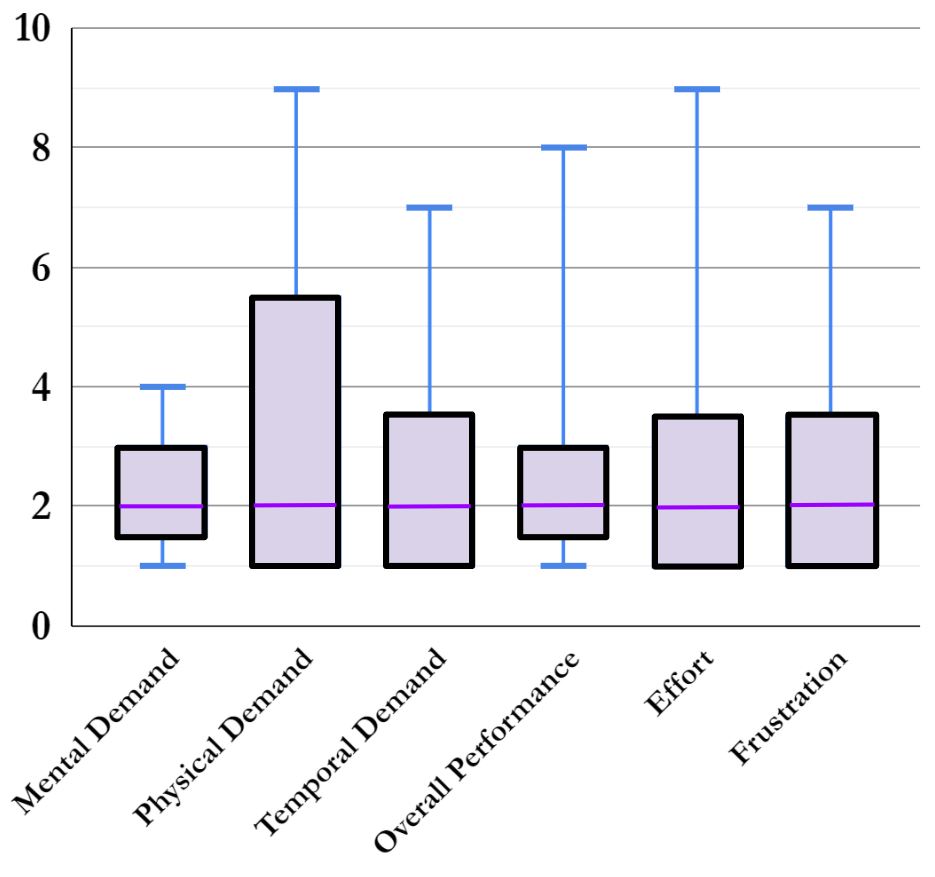}
  \caption{NASA TLX rating results for the six sub-clases during the operation of the robot by the CobotTouch interface.}
  \label{fig:Nasa TLX}
  \vspace{-0.3cm}
\end{figure}

\begin{table}[tb]
\caption{Average NASA TLX Rating}
  \label{table_NASA}
\centering
\begin{tabular}{|l|c|}
\hline
\multicolumn{1}{|c|}{} & CobotTouch System \\ \hline
Mental Demand          & 1.33              \\ \hline
Physical   Demand      & 2.08              \\ \hline
Temporal   Demand      & 1.58              \\ \hline
Performance            & 1.67              \\ \hline
Effort                 & 1.75              \\ \hline
Frustration            & 0.92              \\ \hline
Geneal TLX Score       & 13                \\ \hline
\end{tabular}
\end{table}

Results also showed that participants had the lowest frustration level of 0.92 compared to other types of demands. Several participants noticed that it was simple to observe the controlling interface and the robotic end-effector simultaneously. The Average time of task performing was 5 minutes 19 sec. The fastest performing was 4 min 1 sec., and the slowest was 6 min 48 sec.

\subsection{Post-Experience Questionnaire}

Generally, the users were very inspired by the CobotTouch system. Some of the participants noticed that they could push the robot only by finger in a desirable direction. 
The comments: 

``I liked the projector interface (CobotTouch), it felt like I was pushing the robotic arm, and it was moving." ``I like how easy the robot can be guided by hand. However, my arm was tired at the end of the experiment; for me, it would be comfortable to control the robot by both hands." ``The haptic feedback helped me to understand better when the rotation of the end-effector started and to stop at the right time."

The main advantage of this interface is that a person can move the robot with one hand, and the second hand is free to feel the haptic feedback. Secondly, it is non-obligatory to switch constantly the view from interface to end-effector. The person can observe them simultaneously. Third, the operator gets kinesthetic haptic feedback from the robot, and he can feel where he is moving the robotic arm.

\section{Conclusion and Future Work}

This study presented CobotTouch, a novel robot control interface with a projected GUI, DNN-based gesture recognition for dexterous end-effector manipulation, and fingertip haptic feedback to render supporting information to the user. Our system consists of a projector, web camera, and the novel 2-finger haptic display that provides tactile stimuli to the index and thumb fingertips. 

We evaluated the pattern recognition for the designed haptic display and evaluated the interface's intuitiveness in the robot control task. The result shows that the device demonstrates high recognition average rate of 75\% for the eight tactile patterns and a low NASA TLX rating of 13 scores on average.

In future work, more haptic patterns will be studied during the manipulation of the robot by the proposed system, and a study to compare with other interfaces will be performed. 

The proposed robot control interface could potentially improve industrial and collaborative robot learning by demonstration programming. An intuitive and immersive interface, coupled with highly sensitive tactile feedback, could support the worker during complicated tasks such as peg in a hole, where the visual channel does not provide sufficient information. For future work, we consider extending the functionality of the projected interface by adding new virtual buttons for speed regulation. Besides, we want to study what control actions should be carried out by gestures and what by interaction with the panel.

\section{Acknowledgments}

The reported study was funded by RFBR according to the research project No. 20-38-90294.

\end{document}